\documentclass[a4paper]{jpconf}
\usepackage{graphicx}
\usepackage[]{latexsym,amsmath,amssymb}
\usepackage[]{epsfig}
\newcommand{\be}{\begin{eqnarray}}
\newcommand{\ee}{\end{eqnarray}}
\newcommand{\ud}{\mathrm{d}}

\newcommand{\lp}{\ell_{\rm p}}
\newcommand{\mpl}{m_{\rm p}}

\renewcommand{\bi}{\hat{b}}

\newcommand{\expec}[1]{\mbox{$\langle #1\rangle$}}

\begin{document}
\title{Gravitational renormalization of quantum field theory:
a ``conservative'' approach}
\author{Roberto Casadio}
\address{Dipartimento di Fisica, Universit\`a di Bologna,
and I.N.F.N.,
Sezione di Bologna, via~Irnerio~46, 40126~Bologna, Italy}
\ead{casadio@bo.infn.it}
%
%
%
%
\begin{abstract}
We propose general guidelines in order to incorporate the geometrical
description of gravity in quantum field theory and address the problem
of UV divergences non-perturbatively.
In our aproach, each virtual particle in a Feynman graph should be
described by a modified propagator and move in the space-time generated
by the other particles in the same graph according to Einstein's
(semiclassical) equations.
\end{abstract}
%
%
%
%
\section{Introduction}
\setcounter{equation}{0}
Pauli, long ago~\cite{pauli}, suggested that gravity could act as a
regulator for the ultraviolet (UV) divergences that plague quantum
field theory (QFT) by providing a natural cut-off at the Planck scale.
Later on, classical divergences in the self-mass of point-like
particles were indeed shown to be cured by gravity~\cite{adm},
and the general idea has since then resurfaced the literature many times
(see, {\em e.g.}, Refs.~\cite{deser,dewitt,salam,ford,woodard,robinson}).
In spite of that, Pauli's ambition has never been fulfilled.
\par
As it happens, QFT is successfully used to describe particle
physics in flat~\cite{peskin} (or curved but still fixed~\cite{birrell})
space-time where standard renormalization techniques allow one 
to obtain testable results, notwithstanding the presence of ubiquitous
singularities stemming from the very foundations of the theory,
that is the causal structure of (free) propagators.
We have thus grown accustomed to the idea that the parameters in
a Lagrangian have no direct physical meaning and infinite contributions
may be subtracted to make sense of mathematically diverging integrals.
The modern approach to renormalization~\cite{wilson} views the
occurrence of such infinities as a measure of our theoretical
ignorance of nature and every Lagrangian should, in turn, be considered
as an effective (low energy) description doomed to fail at some UV~energy
scale $\Lambda$~\cite{weinberg97}.
Moreover, gravitational corrections to the Standard Model amplitudes
to a given order in the (inverse of the) Planck mass $\mpl$ are
negligibly small at experimentally accessible
energies~\cite{donoghue}.
These facts briefly elucidate the main theoretical reason that makes
it so difficult to use gravity as a regulator:
if it is to provide a natural solution to the problem of UV~divergences,
gravity must be treated non-perturbatively~\cite{woodard}.
\par
In the QFT community, gravity is mainly viewed as a spin-2~field
which also happens to describe distances and angles (to some extent).
As such, the most advanced strategy to deal with it is the background
field method for functional integrals~\cite{bfm,peskin},
according to which one expands the Einstein-Hilbert Lagrangian
(or a generalisation thereof) around all of the fields' classical
values, including the classical background metric.
The latter is reserved the role of defining the causal structure
of space-time, whereas the quantum mechanical part yields the
graviton propagator and matter couplings (of order $\mpl^{-2}$).
The effect of gravity on matter fields can then be analysed
perturbatively by computing the relevant Feynman
graphs~\cite{veltman}.
A notorious consequence of this approach is that, by simple
power counting, pure gravity is seen to be non-renormalizable,
a ``text-book'' statement~\cite{shomer}, yet occasionally debated.
For example, Ref.~\cite{woodard} suggested that perturbative
expansions are usually performed in the wrong variables and that
Einstein gravity would appear manifestly renormalizable if one
were able to resum logarithmic-like series~\cite{salam}.
In the physically more interesting case with matter,
non-perturbative results can be obtained in just a very few
cases, one of particular interest being the correction to the
self-mass of a scalar particle, which becomes finite once all
ladder-like graphs containing gravitons are added~\cite{dewitt}.
A remarkable approach was developed in Refs.~\cite{reuter98},
in which a tree-level effective action for gravity at the energy
scale $\mu$ is derived within the background field method but
without specifying the background metric {\em a priori\/}.
The latter is instead, {\em a posteriori\/} and self-consistently,
equated to the quantum expectation value determined by the effective
action at that scale.
This method does not involve cumbersome loop contributions and hints
that gravity might be {\em non-perturbatively\/}
renormalizable~\cite{reuter08}, with a non-Gaussian UV~fixed point,
thus realising the {\em asymptotic safety\/} conjectured several
decades ago by Weinberg~\cite{weinberg}.
\par
Based on the idea that QFT is an effective approach~\cite{weinberg97},
different attempts have taken a shortcut and addressed the effects
of gravity on the propagation of matter field directly,
{\em e.g.}, by employing modified dispersion relations or
uncertainty principles at very high ({\em trans-Planckian\/}) energy~\cite{unruh}.
Some works have postulated such modifications, whereas others
tried to derive them from (effective) descriptions of quantum
gravity (see, {\em e.g.\/}, Refs.~\cite{maggiore}).
It is in fact common wisdom that, for energies of the order
of $\mpl$ or larger, the machinery of QFT fails and one will
need a more fundamental quantum theory of gravity, such as
String Theory~\cite{string} or Loop Quantum Gravity~\cite{loop}. 
Quite interestingly, both approaches hint at space-time
non-commutativity~\cite{szabo} as an effective implementation
of gravity as a regulator, with the scale of non-commutativity
of the order of the Planck length $\lp$.
A new feature which, in turn, follows from space-time
non-commutativity is the IR/UV~mixing, whereby physics in the
infrared (IR) is affected by UV~quantities~\cite{IRUV}.
This feature gives us hope of probing (indirectly) such an
extreme energy realm in future experiments or even using available
data of very large scale (cosmological) structures.
\par
In Ref.~\cite{GR}, we proposed yet a different strategy
to incorporate gravity in the body of QFT.
Instead of proposing a new, or relying on an available,
fundamental theory of quantum gravity, we tried to define
modified propagators in a very ``conservative'' (minimal)
way inspired by the simple semiclassical perspective
in which gravity is described by Einstein's geometrical theory
and matter by perturbative QFT.
In this approach, gravity is therefore not viewed as a spin-2
field, but rather as the causal structure of space-time
(or the manifestation thereof) at all loops in QFT, a property
the background field method instead reserves to the classical
part of the metric only.
The modified propagators for matter fields should therefore
take into consideration the presence of each and every source,
classical or virtual, in a given process mathematically described
by Feynman's diagrams.
Of course, philosophical perspectives aside, the relevant question
is whether this idea leads to different (or the same) phenomenological
predictions with respect to the other approaches to UV physics
currently available.
However, we are in a fairly premature stage to assess that.
In fact, even realising the relatively simple guidelines which
we review here poses serious technical problems, and a very
preliminary attempt, based on several further working assumptions,
can also be found in the second part of Ref.~\cite{GR}. 
\par
We shall use units with $c=\hbar=1$ and the Newton constant
$G=\lp/\mpl$.
\section{Geometrical gravity in QFT}
\setcounter{equation}{0}
\label{semi}
In order to make contact with the physics,
let us note that one needs to consider two basic energy scales,
one related to phenomenology and one of theoretical origin,
namely:
\begin{description}
\item[a)]
the highest energy presently available in experiments, say
$E_{\rm exp}\simeq 1\,$TeV, and
\item[b)]
the Planck energy $\mpl\simeq 10^{16}\,$TeV.
\end{description}
It is well assessed that, for energies up to $E_{\rm exp}$,
the Standard Model of particle physics (without gravity) and
renormalization techniques yield results in very good agreement
with the data.
Further, finite, albeit experimentally negligible, quantum
gravitational corrections can be obtained by employing the
effective QFT approach~\cite{donoghue}
(which also yields some -- but not all -- of the general
relativistic corrections to the Newtonian potential).
At the opposite end of the spectrum, for energies of the order
of $\mpl$ or larger, QFT presumably breaks down and one needs a
new quantum theory which includes gravity in a fundamental manner,
like String Theory~\cite{string} or Loop Quantum
Gravity~\cite{loop}.
\par
In any case, we expect that gravitational corrections to QFT
amplitudes play an increasingly important role for larger and
larger energy scale $\mu>E_{\rm exp}$, and that it should be possible
to describe such effects in perturbative QFT directly
(at least in the regime $E_{\rm exp}\lesssim\mu\lesssim \mpl$).
We call this window the realm of ``semiclassical gravity'',
and that is the range where our proposal is more likely to shed
some new light~\footnote{We actually attempted at pushing our
predictions even further and address the very problem of UV
divergences in Ref.~\cite{GR}.}.
\subsection{Semiclassical gravity}
At intermediate energies $E_{\rm exp}\lesssim\mu\ll \mpl$,
we expect that a semiclassical picture holds in which the
space-time can be reliably described as a classical manifold
with a metric tensor $g_{\alpha\beta}$ that responds to the
presence of (quantum) matter sources according to the
well-known equation~\cite{birrell}
\be
R_{\alpha\beta}-\frac{1}{2}\,R\,g_{\alpha\beta}
=\frac{\lp}{\mpl}\,\expec{\hat T_{\alpha\beta}}
\ ,
\label{E}
\ee
where $R_{\alpha\beta}$ ($R$) is the Ricci tensor (scalar)
and $\expec{\hat T_{\alpha\beta}}$ the expectation value of
the matter stress tensor obtained from QFT on that background.
All the same, if one takes Eq.~\eqref{E} at face value,
the way perturbative terms are computed in QFT appears
questionable, since loops of virtual particles are included
in Feyman's diagrams whose four-momentum $|k^2|=|k_\alpha\,k^\alpha|$
formally goes all the way to infinity ({\em i.e.}, to $\mpl$
and beyond) but are still described by the (free) propagators
computed on a fixed (possibly flat) background.
\par
\begin{figure}
\centering
$k$
\\
$p_2$
\hspace{-0.5cm}
\raisebox{3.0cm}{$p_1$}
\epsfxsize=10cm
\epsfbox{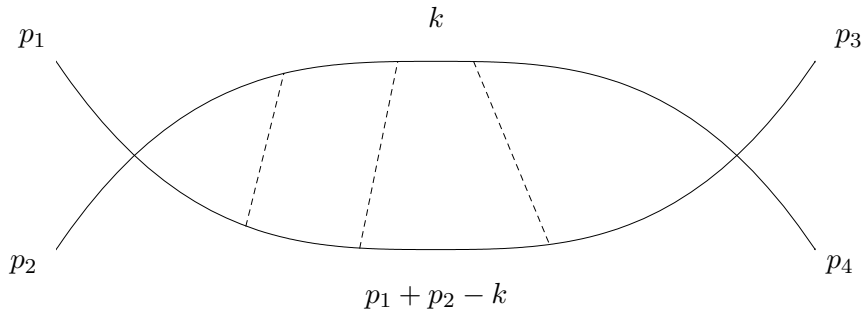}
$p_4$
\hspace{-0.5cm}
\raisebox{3.0cm}{$p_3$}
\\
$p_1+p_2-k$
\caption{Usual one-loop correction to the four-point function
in $\lambda\,\phi^4$ (solid lines) and graviton exchanges
(dashed lines).
\label{loop}}
\end{figure}
For example, let us consider the graph for scalar particles with self-interaction $\lambda\,\phi^4$, represented by solid lines
in Fig.~\ref{loop}, which is a pictorial representation of
the integral
\be
\Gamma^{(4)}(p)
\simeq
\int
\frac{k^3\,\ud k}{(2\,\pi)^4}\,
\tilde G_{\rm F}(k)\,\tilde G_{\rm F}(p-k)
\ ,
\label{G4}
\ee
where $\tilde G_{\rm F}$ is the momenutm-space Feynman propagator
in four dimensions,
\be
\tilde G_{\rm F}(p)
=
\frac{1}{p^2+i\,\epsilon}
\ .
\label{GF}
\ee
Although the external momenta $p_i$ ($i=1,\ldots,4$) are taken
within the range of experiments
(that is, $|p_i^2|\lesssim E_{\rm exp}^2$ in the laboratory frame),
the two virtual particles in the loop have unconstrained
momenta $k$ and $p_1+p_2-k$, respectively.
One might therefore wonder if it is at all consistent
to describe those two particles using the above flat-space
propagator.
The common QFT approach to this problem would result in adding
gravity in the form of graviton exchanges (the dashed lines in
Fig.~\ref{loop})
and estimate deviations from purely flat-space results.
This procedure is however likely to miss non-perturbative
contributions that the UV physics might induce into the IR.
For sure, it will not render finite diverging integrals, such as
the one in Eq.~\eqref{G4}, unless one is able to resum an infinite
number of perturbative terms.
\par
The interplay among propagators, UV~divergences and the causal structure
of space-time can be better appreciated by noting that, in any approach
in which the space-time structure is a fixed background,
the short distance behaviour of QFT (in four dimensions) is described
by the Hadamard form of the propagators~\cite{ford}
\be
G(x,x')=
\frac{U(x,x')}{\sigma}
+V(x,x')\,\ln(\sigma)
+W(x,x')
\ ,
\ee
where $U$, $V$ and $W$ are regular functions and $2\,\sigma$ is the square
of the geodesic distance between $x$ and $x'$.
For instance, in Minkowski space-time, one has
\be
2\,\sigma=(x-x')^2
\ ,
\label{sigma}
\ee
and the propagator contains divergences for $\sigma\to 0$
({\em i.e.}, along the light cone and for $x\to x'$).
Calculations based on the use of propagators in QFT therefore
(implicitly) rely on the formalism of distribution theory and
UV~divergences appear as a consequence when one tries to compute
(mathematically) ill-defined quantities, such as the four-point
function in Eq.~\eqref{G4}.
One can devise mathematical workarounds for this problem, but
what matters here is that, if only the relation~\eqref{sigma}
is modified (like in QFT on a curved space-time), the divergences
for $\sigma\to 0$ will remain.
Nonetheless, a few partial results suggest that deeper
modifications of the causal structure might occur at the quantum level.
For example, it was shown that the divergence on the light-cone
disappears (with a smearing at large momenta of the form considered
in Ref.~\cite{spallucci}) if graviton fluctuations are in a coherent
state~\cite{ford}~\footnote{With the further inclusion of negative norm
states, all UV~divergences were claimed to be cured in
Ref.~\cite{rouhani}.}.
\par
It seems appropriate to us to tackle this problem by pushing further
the validity of  the semiclassical Einstein equations.
We shall hence assume that virtual particles propagate in a background
compatible with Eq.~\eqref{E} at the scale $\mu\sim k=\sqrt{|k^2|}$
and their propagators be correspondingly adjusted~\cite{deser}.
As we mentioned before, our underlying viewpoint is not that gravity
is just a field (although with a very complicated dynamics),
but the geometrical perspective according to which gravity {\em is\/}
the space-time and, in particular, the causal structure obeyed by all
(other) fields.
Let us remark again that this view is partly incorporated
into the background field method, whereby the metric is split into
two parts,
\be
g_{\mu\nu}=\eta_{mu\nu}+h_{\mu\nu}
\ .
\ee
The classical part $\eta_{\mu\nu}$ possesses the expected symmetries
of General Relativity and determines the causal structure for all
(other) classical and quantum fields, whereas $h_{\mu\nu}$ is just
another quantum field which acts on the matter fields via usual
(although complicated) interaction terms, hence in a non-geometrical
way.
One could actually view our approach as a step backward,
since the gravitational field is not explicitly quantised~\cite{ken}
(there is no analogue of the above $h_{\mu\nu}$),
and it is in fact not even defined separately ({\em i.e.},~in the absence
of matter~\footnote{This is somewhat reminiscent of the
``relational mechanics'' approach to gravity (see, {\em e.g.\/},
Ref.~\cite{anderson08} and References therein).}).
\subsection{Gravity in propagators and transition amplitudes}
\label{set}
We can now formulate the basic prescriptions for defining
a ``gravitationally renormalised'' QFT:
\begin{description}
\item[A1)]
perturbative QFT defined by Feynman diagrams is a viable approach
to particle physics for energies $\mu$ below a cut-off
$\Lambda\gg E_{\rm exp}$;
\item[A2)]
in a (one-particle irreducible) Feynman diagram with $N$~internal lines,
each virtual particle is described by a Feynman propagator
\be
G(x,y)=G_{\{x_i\}}^{(\Lambda)}(x,y)
\label{Gx}
\ee
corresponding to the space-time generated by the other $N-1$ virtual
particles in the same graph with coordinate positions $x_i$
($i=1,\ldots,N-1$) and constrained according to~{\bf A1};
\item[A3)]
Standard Model results are recovered at low energy,
$\mu\lesssim E_{\rm exp}\ll\mpl$.
\end{description}
Several comments on the above guidelines are in order.
First of all, we explicitly introduced a cut-off in {\bf A1},
having in mind that our approach is not meant to be the final theory
of everything, but should rather be regarded as a computational recipe.
A second, essential, simplification was introduced in {\bf A2}, in that
each virtual particle is treated like a test particle in the space-time
generated by the other particles, its own gravitational backreaction
thus being neglected~\footnote{Let us note in passing that this
somewhat parallels a perturbative result of non-commutative QFT,
according to which there is no tree-level correction to the commutative
case~\cite{szabo}.}.
Another consequence of {\bf A2} is that integration over positions
inside loops can now be viewed as also purporting a (quantum mechanical) superposition of (virtual) metrics, and there is hope that this can smear
the usual divergences of~\eqref{GF} out (as was shown in Ref.~\cite{ford}
for particular gravitational states).
It also should not go unnoticed that we did not mention a Lagrangian
(or action) from which the modified propagators satisfying {\bf A2}
could be obtained.
In this respect, our proposal follows the philosophy of Ref.~\cite{veltman},
which gives the Lagrangian a secondary role with respect to Feynman's
rules for computing perturbative amplitudes.
However, the symmetries of a system are far more readable
if a Lagrangian is available~\cite{weinberg97} and it would be interesting
to find out whether an action principle can be devised to streamline the
derivation and show which symmetries are preserved or broken.
The latter kind of analysis can also be performed perturbatively, although,
as is well known for the Slavnov-Taylor identities of (non-Abelian)
Yang-Mills theory, that task requires a lot more effort.
A final observation is that the Standard Model of particle physics
(without gravity) is a rigid theory and it is very likely that a generic
modification of the sort we are proposing here has hazardous effects in the
range of presently available data, thus compromising {\bf A3}.
One should therefore check very carefully that none of the assessed
predictions of the Standard Model is lost in our approach.
\par
One cannot ignore the technical fact that the $N$-body problem in
General Relativity is extremely complicated, to say the least,
already for $N=2$.
To the general guidelines, we therefore add two working assumptions:
\begin{description}
\item[W1)]
starting from the coordinate-space propagator in Eq.~\eqref{Gx},
it is possible to define a momentum-space propagator
$\tilde G_{\{k_i\}}(p)$;
\item[W2)]
one can approximate the momentum-space propagator for each virtual
particle
\be
\tilde G_{\{k_i\}}^{(\Lambda)}(p)\simeq \tilde G_{q}^{(\Lambda)}(p)
\ ,
\label{Ggr}
\ee
where $q\simeq \sqrt{|\sum k_i|^2}$ is the total momentum of the
remaining $N-1$ particles.
\end{description}
The latter is a ``mean field'' assumption devised to deal with graphs
containing more than two virtual particles.
The approximate equality in Eq.~\eqref{Ggr} may thus be replaced
with other expressions of choice, the key point being that the
problem reduces to studying the propagator for a test particle in
a background generated by an ``average'' source.
\par
For example, the ``gravitationally renormalized'' analogue of the
four-point amplitude in Eq.~\eqref{G4} is obtained by replacing
each particle's propagator with the new expression~\eqref{Ggr},
\be
\Gamma^{(4)}_{\rm GR}(p;\Lambda)
\simeq
\int^\Lambda
\frac{k^3\,\ud k}{(2\,\pi)^4}\,
\tilde G_{(p-k)}^{(\Lambda)}(k)\,\tilde G_{(k)}^{(\Lambda)}(p-k)
\ .
\label{G4gr}
\ee
Provided $\tilde G_{(k)}(q)$ falls off fast enough at large $k$,
one can therefore hope to obtain finite transition amplitudes
even in the limit $\Lambda\to\infty$.
Of course, in order to obtain explicit expressions, one first needs
to solve for the geometry produced by virtual particles and then
obtain the momentum-space form of the propagator (an
early attempt can be found in the second part of Ref.~\cite{GR}).
\section{Final remarks}
\setcounter{equation}{0}
Inspired by the observation that a semiclassical description of gravity
should be possible in processes that involve energies below the Planck scale,
we formulated general guidelines that can be employed to adjust QFT
in order to include gravitational contributions.
Such guidelines were listed in the form of general prescriptions
({\bf A1}-{\bf A3}) and more specific working assumptions
({\bf W1} and {\bf W2}) that formalise our approach to include gravity
within the Standard Model of particle physics in a geometrical and
non-perturbative way.
All of them are of course debatable and subject to possible refinements.
\par
Preliminary results, reported in Ref.~\cite{GR}, showed that one might
indeed expect significant UV modifications from the dependence of the
propagators on the momenta of virtual particles.
However, further working assumptions were used therein, whose impact
must be clarified.
And, of course, a realistic QFT should be analysed before the final
word can be spoken on UV divergences and that old idea of Pauli.
\section*{References}
\end{document}